\documentclass[aps,10pt,prd,a4paper,twocolumn,notitlepage,nofootinbib,showpacs]{revtex4-1}
\usepackage{amsmath}
\usepackage{graphicx}
\usepackage{color}
\usepackage[normalem]{ulem} %PJ

\allowdisplaybreaks

\newcommand{\be}{\begin{equation}}
\newcommand{\ee}{\end{equation}}

\newcommand{\pa}{\partial}

\newcommand{\bea}{\begin{eqnarray}}
\newcommand{\eea}{\end{eqnarray}}
\newcommand{\ben}{\begin{eqnarray*}}
\newcommand{\een}{\end{eqnarray*}}

\begin{document}
 
\title{General-relativistic rotation laws in  fluid tori
around spinning black holes }

\author{Wojciech Kulczycki}
\author{Edward Malec}
\affiliation{Instytut Fizyki im.~Mariana Smoluchowskiego, Uniwersytet Jagiello\'nski, {\L}ojasiewicza 11, 30-348 Krak\'{o}w, Poland} 

\begin{abstract}
	We  obtain   rotation laws for  axially symmetric, selfgravitating and stationary fluids around spinning black holes. They reduce --- in the Newtonian limit   ---  to  monomial  rotation curves.  For spinless black hole, one obtains in the first post-Newtonian (1PN) approximation   the hitherto  known results, that can be interpreted as  the  geometric dragging  and material antidragging. We find new 1PN   effects,   that are due to  spins of black holes.

\end{abstract}


\maketitle 

\section{Introduction}

In axially symmetric, stationary Newtonian hydrodynamic configurations, consisting of a massive central body and a (barotropic) fluid toroid  circulating around a fixed axis,   the  angular momentum per unit mass  $j$ can be any function of $r$. Here $r$ is the distance from the rotation axis. This is known as the integrability condition of Wavre and Poincare` \cite{Tassoul}. Other restrictions are due to a stability condition (see a discussion in \cite{Tassoul}). In particular, the angular velocity can have the monomial form $\Omega_0=w_0/r^{2\over 1-\delta} $, where    $-\infty  < \delta  \le 0$ and $w_0$ is a real number. Prescribing  the angular velocity (giving the rotation curve or the
rotation law, in the specialists' terminology) turns the Euler-Poisson equations of stationary hydrodynamics into a closed system, that can be analysed numerically.

An integrability condition exists also, in axially symmetric stationary (barotropic) fluids, for general relativistic systems; in such a case the angular momentum  per unit mass $j$   is any function $j=j(\Omega )$ \cite{Bardeen}; here $\Omega $ is the angular velocity. As before, the analytic problem becomes closed once a rotation law $j(\Omega )$ is specified. The main challenge is to find a rotation law that is realistic and at the same time  solvable numerically. Uniformly (rigidly) rotating    gaseous disks in general-relativistic hydrodynamics have  been  discussed in \cite{Bardeen,Ipser}.   A more realistic  angular velocity profile has been  studied since   1980's --- in the context of rotating stars ---   with the angular momentum density being  a linear function of the frequency  \cite{komatsu,nishida_eriguchi,nishida1}. We should mention a later investigation of the rigid rotation by members of  the Jena group \cite{NM,AP,MAKNP}.    Quite recently  various differential (nonlinear)  rotation laws have been analysed \cite{GYE,UTG,UTB,TUS}. 

The present paper is a  continuation of the work \cite{MM} which  gives  
 rotation laws for polytropic fluids in motion around spinless black holes. In what follows,  we find a generalized rotation curve, that includes also spinning black holes. Its special case  is the recently found general-relativistic keplerian rotation, that appeared to be solvable numerically \cite{KKMMOP, KKMMOP1}.
It appears that in the non-relativistic limit one gets a monomial angular velocity $\Omega_0=w_0/r^{\frac{2}{\delta -1}} $.     

\section{Hydrodynamical equations}
Below we write down, after     \cite{komatsu}, the   hydrodynamic equations.
Einstein equations, with the signature of the metric $(-,+,+,+)$, read
\begin{equation}
R_{\mu \nu} -g_{\mu \nu }\frac{R}{2} = 8\pi \frac{G}{c^4}T_{\mu \nu }.
\label{ee1}
\end{equation}
  $T_{\mu \nu }$  denotes the stress-momentum tensor.  The metric is \emph{stationary}, and it is given by

\begin{align}
\label{metric}
d s^2 &=  -  e^{\frac{2\nu }{c^2} }(d x^0)^2
+r^2   e^{\frac{2\beta }{c^2} } \left( d \phi  -  \frac{\omega }{c^3}  d x^0\right)^2
\nonumber\\&\qquad
+e^{\frac{2 \alpha }{c^2} }   \left( dr^2 +   dz^2\right)    .
\end{align}
Here $c$ is the speed of light,  $x^0 =ct$ is the   time coordinate, and $r$, $z$, $\phi$ are cylindrical  coordinates.  
 We assume axial symmetry and use  the stress-momentum tensor  
\be 
\label{emD}
T^{\alpha\beta} = \rho (c^2+h)u^\alpha u^\beta + p g^{\alpha\beta},
\ee
where $\rho$ is the baryonic rest-mass density of a fluid, $h$ is the  specific enthalpy,
and $p$ is the  pressure.  
The 4-velocity  {$u^\alpha  $}  is normalized, $g_{\alpha\beta}u^\alpha u^\beta=-1$.
 The coordinate (angular) three-velocity reads ${\vec v}= \Omega \partial_\phi $, where $\Omega = u^\phi /u^t$.

We  take the  polytropic equation of state  $p(\rho ,S) = K(S) \rho^\gamma$,
where $S$ is the specific entropy of fluid. Then one finds $h(\rho ,S) = K(S) \frac{\gamma}{\gamma-1}\rho^{\gamma-1}$. The entropy is  constant.

The square of the linear velocity in the locally stationary coordinate system is given by
\be
V^2=r^2 \left( \Omega -\frac{\omega }{c^2}\right)^2 e^{2\left( \beta - \nu \right)/c^2}.
\label{vel}
\ee
The potentials $\alpha$, $\beta$, $\nu$, and $\omega$ satisfy  equations that have been found by Komatsu, Eriguchi and Hachisu \cite{komatsu}.  They constitute an over-determined, but consistent,  set of equations. The general-relativistic Euler equations are solvable, assuming  an  integrability condition --- that    the angular momentum per unit mass,
\be
j  =  u_\phi u^t= \frac{V^2}{\left( \Omega -\frac{\omega }{c^2}\right) \left( 1-\frac{V^2}{c^2}\right) },
\label{ampum}
\ee
depends only on the angular velocity $\Omega $: $ j\equiv j(\Omega )$ \cite{Bardeen}. In such a case  the Euler equations reduce to  a general-relativistic integro-algebraic  equation, that    embodies  the hydrodynamic information carried by the continuity equations  $\nabla_\mu T^{\mu \nu }=0$ and the baryonic mass conservation
$\nabla_\mu \left( \rho u^\mu \right) =0$.  It is given by the expression  
\be
\label{grBernoulli}
\ln \left( 1+\frac{h}{c^2}\right) +\frac{\nu }{c^2} +\frac{1}{2}\ln \left( 1-\frac{V^2}{c^2}\right) +\frac{1}{c^2}\int d\Omega j(\Omega ) =C.
\ee
The quantity on the left hand side of (\ref{grBernoulli}) is constant not only along the flow of fluid (as in the case of the Bernoulli equation), but within the whole fluid volume. Nevertheless, we shall sometimes use the name ``Bernoulli equation", when referring to (\ref{grBernoulli}).

\section{New rotation laws}

The   new family of rotation laws reads
\begin{eqnarray}
\label{momentum}
j(\Omega ) &\equiv &-\frac{c^2}{1-3\delta}\frac{d}{d\Omega} \ln \left(  1-  (\frac{a\Omega}{c})^2\right. \nonumber\\
&& \left.- \frac{ \kappa  }{   c^2 }  w^{1- \delta }\Omega^{1+\delta }(1-\frac{a\Omega}{c})^{1-\delta} \right) .
\end{eqnarray}
Here $J$ and  $a=J/(Mc)$ are the angular momentum and the spin parameter of the central black hole, respectively.    $\delta $ is a free  parameter and $\kappa =\frac{1-3\delta }{1+\delta }$.
 
 In the Newtonian limit $c\rightarrow \infty $   one arrives at 
\be
\label{zeroth_rotation_law}
\Omega_0 = \frac{w_0}{r^\frac{2}{1- \delta }},
\ee
where $\Omega_0= \lim_{ c\rightarrow \infty}\Omega $ and $  w_0=\lim_{ c\rightarrow \infty} w$.  $\delta $  can be  freely chosen within the interval  $(-\infty,  0]$, with the reservation that $\delta \neq -1$ (this condition can be removed at least for tori circulating around  spinless black holes \cite{KMM}). The two limiting cases $\delta  =0$  and $ \delta =-\infty $ correspond to the constant angular momentum per unit mass  ($\Omega_0 =w_0/r^2$ and $j=w_0$)  and the rigid rotation ($\Omega_0 =w_0$ in the spinless case), respectively.
 
We would like to stress that the   constant $w $ is any real number, but    it is not free, excepting for test-like tori; for massive tori  the value of $w$ actually is a part of the solution \cite{KKMMOP, KKMMOP1}.       The choice of $ \delta =-1/3$ yields the generalization of the Keplerian rotation that has been investigated in \cite{KKMMOP, KKMMOP1}. In the Newtonian limit we have, for massless disks, $w^2_0=GM_\mathrm{c}$ and $\Omega_0=\sqrt{GM_\mathrm{c}}/r^{3/2}$, where $M_\mathrm{c}$ is the central mass \cite{MMP}.
 
Notice that (\ref{momentum}) coincides with the formerly found rotation curve \cite{MM},   if the black-hole spin parameter vanishes, $a=0$.  The present rotation law is valid in the same interval of $\delta $, but it applies both to spinning and spinless black holes.
  
The next two sections are dedicated to the description of a reasoning that yields  
(\ref{momentum}).

\section{ Angular velocity via post-Newtonian expansion: 1PN corrections}

In the first post-Newtonian  (1PN) approximation  one chooses  the metric exponents
$\alpha =\beta =-\nu =-U$ with  $|U|\ll c^2$ \cite{BDS}. Define $A_\phi \equiv r^2\omega $. The spatial part of the obtained metric, 
\begin{align}
\label{metric1}
d s^2 &=  -  \left(1+\frac{2U}{c^2} +\frac{2U^2}{c^4}\right) (d x^0)^2
-2  c^{-3} A_\phi d x^0 d \phi \ 
\nonumber\\&\qquad
+\left(1-\frac{2U}{c^2}  \right)  \left( d r^2 + d z^2 + r^2 d \phi^2\right),
\end{align} 
becomes now conformally flat.  One needs to solve the full system of 1PN approximations of Einstein and hydrodynamical equations, in order to find numerical values of 1PN corrections to the angular velocity.  We want, however, to find only the functional form of corrections to the angular velocity; in this case it suffices to consider the hydrodynamic equations.

Let   $\rho$ and   $h$ be the  mass density and  the specific enthalpy of a fluid, respectively.
As in \cite{BDS,JMMP}, we split  $\rho$,    $h$ and the potential $U  $  into their Newtonian (denoted by the subscript $`0'$) and 1PN (denoted by the subscript $`1'$) parts:
\begin{eqnarray}
&&\rho = \rho_0 + \frac{ \rho_1}{c^2}, 
\nonumber\\ 
&& h=h_0+\frac{h_1}{c^2},
\nonumber\\  
&&U = U_0 + \frac{U_1}{c^2}.
\label{density_rotation}
\end{eqnarray}
The angular velocity is decomposed as
\begin{equation}
\Omega  = \frac{w}{r^\frac{2}{1- \delta }} +\frac{\Omega_1}{c^2}.
\label{omegaw}
\end{equation}
Notice that, up to the 1PN order,   
\begin{equation}
\label{enthalpy}
\frac{1}{\rho} \partial_i p = \partial_i h_0 + c^{-2} \partial_ih_1
 {+ \mathcal{O}(c^{-4})},
\end{equation}
where the {1PN} correction $h_1$ to the specific enthalpy can be written as $h_1 = \frac{dh_0}{d\rho_0} \rho_1$. For the polytropic equation of state this gives $h_1= \left( \gamma -1 \right) h_0 \rho_1 / \rho_0$.

The expression for the angular velocity can be written in yet another form
 \begin{equation}
\label{avgeo}
\Omega = \frac{w}{r_\mathrm{c}^{2/(1-\delta)}}    -\frac{2}{c^2(1 -  \delta )} \Omega_0 U_0+\frac{\Omega_1}{c^2};
\end{equation} 
notice that in the Newtonian gauge that is assumed in (\ref{metric1}), the geometric (circumferential) distance to the rotation axis   is given by $r_\mathrm{c}=r(1-U_0/c^2) $, in the leading order.   
 
The  Newtonian hydrodynamic equation reads \cite{JMMP}
\be
\label{0Bernoulli}
h_0+U_0+\frac{1- \delta }{2(1+ \delta )} \Omega_0^2r^2  = C_0,
\ee
where $C_0$ is a constant that  can be interpreted as the binding energy per unit mass.
This is supplemented by the Poisson equation for the gravitational potential 
\be
\label{DeltaU0}
\Delta U_0 = 4\pi G \left( M_\mathrm{c}\delta(\mathbf{x})   +\rho_0\right),
\ee
where $\Delta$ denotes the flat Laplacian. $M_\mathrm{c}$ is the central mass;  $U_0$ is a superposition of the central potential $-GM_\mathrm{c}/\sqrt{r^2+z^2}$ and the potential generated by a torus.

The 1PN equation for the metric function $U_1$ reads \cite{JMMP}
\begin{eqnarray}
\label{U0U1_bis}
\label{DeltaU1}
\Delta U_1 &=& 4\pi G \left( M_{\textrm{c}} U_0 \delta(\mathbf{x}) + \rho_1 + 2p_0 \right. \nonumber \\ 
&& \left. + \rho_0(h_0-2U_0+2 r^2\Omega_0^2) \right).
\end{eqnarray}
The component $A_\phi $ of the shift vector  satisfies the following equation \cite{KMMPX}
\be
\label{Afi}
\Delta A_\phi -2\frac{\pa_rA_\phi }{r}= -16 \pi  Gr^2 \rho_0 \Omega_0 .
\ee
 The 1PN hydrodynamical equations shall be given in terms of a scalar function $\Psi $, that is defined as follows
\begin{eqnarray}
\label{Psi}
\Psi &=& -h_1 - U_1 -\Omega_0 A_\phi + 2 r^2 (\Omega_0)^2 h_0 - \nonumber\\ && {3\over 2} h^2_0 - 4 h_0 U_0 - 2 U_0^2
- \int \mathrm{d}r\, r^3(\Omega_0)^4.
\end{eqnarray}
The 1PN hydrodynamic equations read now, 
\begin{subequations}
\label{euler1gradPN}
\begin{align}
\pa_z\Psi &= 0,
\\[1ex]
\partial_r \Psi + 2r \Omega_0 \Omega_1 + A_\phi \pa_r \Omega_0
-2r^2\pa_r (\Omega_0)^2  h_0 &= 0,
\end{align}
\end{subequations}
There emerges a consistency condition, that yields the form of $ \Omega_1$.
 Namely, differentiating the first and second equation in (\ref{euler1gradPN}) with respect $r $ and $z$, respectively, and subtracting the obtained equations, one arrives at 
\begin{equation}
\label{euler1aaPN}
2r \Omega_0\pa_z\Omega_1 +  (\pa_r \Omega_0)(\pa_z A_\phi) 
 -2r^2\pa_r (\Omega_0)^2  \pa_z h_0 =0.
\end{equation}
This constraint is resolved by 
\begin{equation}
\label{constraint_solution}
\Omega_1 = -{ A_\phi \over 2r \Omega_0} \pa_r \Omega_0
+ 2rh_0 \pa_r\Omega_0 + \chi (r),
\end{equation}
as can be checked by direct inspection. The  function $\chi (r)$ is not determined, at this stage;  in fact, it seems to be arbitrary.
This is obviously not acceptable, because the angular velocity is a physical observable.

The full expression for the angular velocity, up to the terms ${ \mathcal{O}(c^{-4})}$, is now given by:
 \begin{eqnarray}
\label{angular velocityc2}
\Omega =\frac{w}{r_\mathrm{c}^{2/(1-\delta)}}    -\frac{2}{c^2(1 -  \delta )} \Omega_0   U_0  +\frac{\Omega_1}{c^2}, 
\end{eqnarray}
where $\Omega_1$ is given by
 \begin{eqnarray}
\label{angular velocityc3}
\Omega_1= \frac{A_\phi}{r^2  \left( 1- \delta \right)}   -  \frac{4}{ (1 -  \delta )} \Omega_0 h_0  +\chi (r).
\end{eqnarray} 
We shall split the function $\chi $ into two parts,
\begin{equation}
\chi = -\frac{2}{ (1 -  \delta )}  \Omega^3_0r^2 + \chi_1 (r,a),
\label{chi}
\end{equation} 
where by assumption   $\chi_1(r,a=0)=0$. Thence
 the angular velocity is given by:
 \begin{eqnarray}
\label{angularvelocityc3}
\Omega &=& \frac{w}{r_\mathrm{c}^{2/(1-\delta)}}    -\frac{2}{c^2(1 -  \delta )} \Omega_0   \left( U_0  + \Omega^2_0r^2\right) +\nonumber\\
&& \frac{A_\phi}{r^2 c^2\left( 1- \delta \right)}   -  \frac{4}{c^2(1 -  \delta )} \Omega_0 h_0  +\frac{\chi_1 (r,a)}{c^2}.
\end{eqnarray} 
The last three terms vanish identically in the case of a massless disk of dust around a Schwarzschild black hole. The rotation law for the disk of dust in a Schwarzschild spacetime  is the same as for a test particle, that is Keplerian (notice that $\delta =-1/3$):  $\Omega = \frac{\sqrt{GM}}{r_\mathrm{c}^{3/2}}$ \cite{Hartle}. Thus the second  term must vanish in the leading order, as it does, in fact.
The term $\chi_1(r,a)$  depends on the angular momentum of the black hole $J=acM$. It should  be  fixed by analyzing the motion of test particles in the Kerr geometry.
Unfortunately,  the spatial forms of the metric for the
Kerr solution in Boyer-Lindquist or Kerr-Schild coordinates, are not conformally flat up to the 1PN order, and the formulae
derived above do not hold. 
 It appears, however, that  the Kerr solution expressed in harmonic coordinates \cite{Kerr_harmonic_coordinates} does possess a   conformally flat spatial metric  in the 1PN approximation (see Appendix).   

One can check by a straightforward calculation, that the angular velocity $\Omega$ of test  particles in a Kerr spacetime, reads as follows at the equatorial plane (again up to the  1PN approximation --- see Appendix),
\begin{eqnarray}
\label{freqkerr}
\Omega  & = & \frac{\sqrt{GM}}{r_{\textrm{c}}^{3/2}}+\nonumber\\
&& c^{-2}\left(\frac{3J^{2}\sqrt{G}}{4M^{3/2}r_{\textrm{c}}^{7/2}}-\frac{GJ}{r_{\textrm{c}}^{3}}\right).
\end{eqnarray}
The function $A_\phi $ --- again at the equatorial plane --- is given by
\begin{equation} 
\label{afi}
A_{\phi} = \frac{2GJ}{ r };
\end{equation}
taking this into account and comparing (\ref{angularvelocityc3}) with (\ref{freqkerr}),
  one obtains for  massless  dust
  \begin{equation}
 \chi_1(r,a)= \left(\frac{3J^{2}\sqrt{G}}{4M^{3/2}r_{\textrm{c}}^{7/2}}-\frac{5GJ}{2r_{\textrm{c}}^{3}}\right).
  \label{end1}
  \end{equation}
  In conclusion,  the 1PN approximation to the Kerr geometry suggests the following form of the     correction   to the angular velocity:
  \begin{eqnarray}
\label{angularvelocityc4}
\Omega &=& \frac{w}{r_\mathrm{c}^{2/(1-\delta)}} -\frac{2}{c^2(1 -  \delta )} \Omega_0 U_0+\frac{\Omega_1}{c^2} = \nonumber\\ &&\frac{w}{r_\mathrm{c}^{2/(1-\delta)}}-\frac{2}{c^2(1 -  \delta )} \Omega_0   \left( U_0  + \Omega_0^2r^2\right) +\nonumber\\
&& \frac{A_\phi}{r^2 c^2\left( 1- \delta \right)}   -  \frac{4}{c^2(1 -  \delta )} \Omega_0 h_0   +\nonumber\\&& 
\lambda_1(\delta)\frac{J^{2}\Omega_0}{c^2M^{2}r^{2}}-\lambda_2(\delta)\frac{J\Omega_0^2}{c^2M} .
\end{eqnarray}   
 The functional dependence of $\lambda_1$ and $\lambda_2$ on the parameter $\delta $ is found in the next section, and it is given below
\begin{eqnarray}
 \label{ll}
 \lambda _1&=&\frac{2}{(1-3\delta)(1-\delta)},\nonumber\\ && \lambda _2= \frac{(2+\delta)}{1+\delta}.
\end{eqnarray}
  In the case of the Kerr geometry
one has $\lambda_1(\delta=-1/3)=3/4$ and 
$\lambda_2 (\delta=-1/3)=5/2$; this agrees with  (\ref{end1}), as it should.

\section{Prescribing rotation laws: angular velocity   of fluids around spinning black holes.}
 
 The equality $j(\Omega )=u_\phi u^0$ allows one --- prescribing a rotation law $j(\Omega )$ ---   to find  the coordinate representation of  the angular velocity $\Omega $.
  Using   (\ref{ampum}) and (\ref{momentum}), one arrives at the equation
  \begin{eqnarray}
\label{momentum1}
&& -\frac{c^2}{1-3\delta}\frac{d}{d\Omega} \nonumber \\
&&\ln \Big(  1-  (\frac{a\Omega}{c})^2-   \frac{ \kappa  }{   c^2 }  w^{1- \delta }\Omega^{1+\delta }(1-\frac{a\Omega}{c})^{1-\delta   }  + \frac{4C_0}{c^2}\Bigr) \nonumber\\
&=& \frac{V^2}{\left( \Omega -\frac{\omega }{c^2}\right) \left( 1-\frac{V^2}{c^2}\right) }.
\end{eqnarray}
 Here the velocity $V^2$  is given by Eq. (\ref{vel}). $C_0$ is the same constant that appears in (\ref{0Bernoulli}).
 
Let us expand   metric functions and the angular velocity
\begin{equation} \Omega =\frac{w}{r^\frac{2}{1- \delta }} +\frac{\tilde\Omega_1 }{c^2}
\label{omegaw1}
\end{equation} 
in powers of $c^{-2}$, as in (\ref{metric1}) and (\ref{density_rotation}). 
 
Then, keeping terms up to $c^{-2}$, replacing $a$ by $J/(cM)$ and using (\ref{omegaw1}),   one gets  (after suitable reordering and  simplification) the left hand side of Eq. (\ref{momentum1}), 
\begin{eqnarray}
&&\frac{w }{r^\frac{2\delta}{1- \delta }}+\frac{1}{c^{2}}\left[ \frac{1-3\delta}{1+\delta} \frac{w^{3 }}{r^{\frac{2(1+2\delta)}{1- \delta }}}+ \delta r^2  \tilde{\Omega}_{1}-4C_0  \frac{w }{r^\frac{2\delta}{1- \delta }}+\right.\nonumber\\ &&\left. \frac{2}{(1-3\delta)}\frac{w }{r^\frac{2}{1- \delta }}\frac{J^{2}}{M^2}- \frac{(1-\delta)(2+\delta)}{1+\delta}  \frac{w^2 }{r^{\frac{2(1+\delta )}{1- \delta }}}\frac{J}{M} \right];
\end{eqnarray}
whereas   the right hand side of Eq. (\ref{momentum1}) yields
\[
r^{2}\frac{w }{r^\frac{2}{1- \delta }}+\frac{1}{c^{2}}\left[r^{4}\frac{w^3 }{r^\frac{6}{1- \delta }}-A_{\phi}-4r^{2}U_{0}\frac{w }{r^\frac{2}{1- \delta }}+r^{2}\tilde{\Omega}_{1}\right].
\]

We shall compare  separately $c^0$ and $c^{-2}$ terms. There appear two terms in the zeroth order, that cancel identically.
The $c^{-2}$ order yields  
\begin{eqnarray}
 0&=&\frac{1-3\delta}{1+\delta} \frac{w^3 }{r^\frac{2(1+2\delta)}{1- \delta }} +\delta r^2\tilde{\Omega}_{1}-r^{4}\frac{w^3}{r^\frac{6}{1- \delta }}+A_{\phi}+\nonumber\\ 
 &&
4r^{2}U_{0}\frac{w}{r^\frac{2}{1- \delta }}-r^{2}\tilde{\Omega}_{1}+\frac{2}{(1-3\delta)}\frac{J^{2}}{M^2}\frac{w }{r^\frac{2}{1- \delta }}-\nonumber\\ 
&&  \frac{(1-\delta)(2+\delta)}{(1+\delta)} \frac{w^2 }{r^\frac{2(1+\delta)}{1- \delta }}\frac{J}{M}-4C_0  \frac{w }{r^\frac{2\delta}{1- \delta }}.
\label{tildeomega}
\end{eqnarray}
 
From the second equation   we get the 1PN correction $\tilde{\Omega}_{1}$,
\begin{eqnarray}
\tilde{\Omega}_{1}&=&   -\frac{2}{1-\delta}r^{2}\Omega_{0}^{3}+ \frac{1}{1-\delta}r^{-2}A_{\phi}+\frac{2}{1+\delta}r^{2}\Omega_{0}^{3}+\nonumber\\&& \frac{4}{1-\delta}U_{0}\Omega_{0}-\frac{4}{1-\delta}C_{0}\Omega_{0}+\nonumber\\&& \frac{2}{(1-3\delta)(1-\delta)}\frac{J^{2}}{r^{2}M^{2}}\Omega_{0} - \frac{(2+\delta)}{1+\delta} \frac{J}{M}  \Omega_{0}^{2}.
\label{tilde}
\end{eqnarray}
Notice that in (\ref{tilde})  the term $\frac{w}{r^{\frac{2}{1-\delta}}}$ is   replaced by $\Omega_{0}$; that is permissible, since this replacement impacts only  terms beginning from  the 2PN order.

The third, fourth and fifth  terms can be replaced  by $ -\frac{4}{1-\delta}h_{0}\Omega_{0} $ --- see (\ref{0Bernoulli}). Finally one arrives at
\begin{eqnarray}
\tilde{\Omega}_{1}&=&  -\frac{2}{1-\delta}r^{2}\Omega_{0}^{3}+\frac{1}{1-\delta}r^{-2}A_{\phi}-\nonumber\\&& \frac{4}{1-\delta}h_{0}\Omega_{0} + \frac{2}{(1-3\delta)(1-\delta)}\frac{J^{2}}{r^{2}M^{2}}\Omega_{0}-\nonumber\\&&\frac{(2+\delta)}{1+\delta} \frac{J}{M}  \Omega_{0}^{2}.
\label{tilde1}
\end{eqnarray}

The comparison of   (\ref{tilde}) with (\ref{angularvelocityc4}) allows us to conclude that
\[
\tilde{\Omega}_{1}=\Omega_{1},
\]
provided that $\lambda_1$ and $\lambda_2$ are as in (\ref{ll}).
The full angular velocity up to the 1PN order is given by (\ref{angularvelocityc4}). The rotation law (\ref{momentum})
can be obtained from the left hand side of (\ref{momentum1}), by setting $C_0=0$ in there. 
 
 There are two known cases, when a solution $\Omega $ of (\ref{momentum1}) is actually an exact solution; both of them correspond to $\delta=-1/3$, $C_0=0$ and a massless disk of dust. 
 It was proven in \cite{MM} that in the Schwarzschild spacetime
 Eq. (\ref{momentum1}) yields $ \Omega =\sqrt{GM}/R^{3/2}$; the angular velocity for circular orbits of particles of dust in a massless disk. Similarly, in the Kerr  spacetime,  one gets from  (\ref{momentum1}) an exact and analytic angular velocity $\Omega $ of a disk of dust \cite{KKMMOP}.

 \section{Discussion}   
 
We can now interpret the meaning of various contributions to the  angular velocity $\Omega $ given by  (\ref{angularvelocityc4}). 
 The first term is   the Newtonian (monomial) rotation law, written as a function of the geometric distance from the rotation axis, in the 1PN order. 
 The second term in (\ref{angularvelocityc4}) vanishes at the plane of symmetry, $z=0$, for circular Keplerian motion of test fluids in the monopole  potential $-GM/\sqrt{r^2+z^2}$.  It depends   both on the contribution of the disk's self-gravity at the plane $z=0$, as well on   the deviation  from the strictly Keplerian motion $\Omega = \sqrt{GM}/r_\mathrm{c}^{3/2}$.
 
 Before continuing, let us define  the notions of dragging and anti-dragging within a a rotating gaseous torus.    We shall say, that a torus is dragged by a particular effect, if the corresponding term in (\ref{angularvelocityc4}) is nonnegative. If the term is negative then we shall refer to the related effect as ``anti-dragging''. 
 
 The third  term is explicitly positive   for the Kerr geometry --- see Eq. (\ref{afi}). It is  also proven to be positive for spinless black holes \cite{JMMP}. Thus   in these two cases a torus is being dragged (forward). The next    term is explicitly negative ---  it reveals the anti-dragging effect that  agrees (for  monomial angular velocities $\Omega_0 =r^{-2/(1- \delta )}w_0$)  with the result obtained earlier in \cite{JMMP,MM}.  The last but one term, proportional  to the square of the black hole spin, is strictly positive, hence it drags gas particles forward. The last term slows gas particles in the comoving case (when the direction of the rotation of the torus agrees with the spin),  and increases their angular velocities if       the torus rotates in the opposite direction to the spin of the black hole.

 Let us remark, that a massless disk made of dust in the Kerr geometry is special --- the first correction term in  (\ref{angularvelocityc4}) and the  specific enthalpy $h_0$ vanish and the disk  is   exposed only to the interaction with  the spin; there are three spin terms, which combine into the   second term in the r.h.s. of Eq. (\ref{freqkerr}). It is easy to check that this term is strictly negative for disks  comoving with angular momentum --- thus we have anti-dragging in the sense defined above. In the case of counter-rotation, the term is strictly positive, hence a massless disk of dust is being dragged.
 
   It is well known,  that a test disk of dust 
 in the Schwarzschild spacetime rotates in the strictly Keplerian way. 
 
 In both cases, that correspond to $\delta =-1/3$ in the rotation law (\ref{momentum}) and $w^2=GM$, we have strict analytic solutions of Einstein and hydrodynamical equations.
 We should add, that the general-relativistic Keplerian law (with $\delta =-1/3$) has been already investigated in \cite{KKMMOP, KKMMOP1}, both for light or heavy fluid disks.
 
 For the  rigid (uniform) rotation, those correction terms $\Omega_1$ that are proportional to $1/(1-\delta )$  do vanish, because  now  $\delta = - \infty$. There remains, however,  one term: $-J\Omega^2_0/M$. That means that the angular velocity jumps from the Newtonian  value $\Omega_0$ to the shifted one --- we stress again: up to the  1PN order --- $ \Omega_0 -J\Omega^2_0/(Mc^2)$, but the rotation is still uniform.

\section{Concluding remarks}

In summary, we have obtained the rotation law    by  the combination of  two different methods. In Sec. IV the 1PN method has been used. In Sec.  V we employed the rotation law and the definition of the angular momentum density $j$ (\ref{momentum1}). In both cases the $c^{-2}$ correction is the same, after fixing the 1PN procedure and some adjustment of coefficients in the rotation law.
The rotation law (\ref{momentum}) is consistent with the 1PN approximation.  

We formulate --- in a reasoning, that can be described as an educated guess --- a family of   non-Keplerian general-relativistic rotation laws, that describe the motion of fluid tori around  spinning black holes. Weak field limit reveals the formerly discovered effects \cite{JMMP, MM} --- dragging and anti-dragging --- and new ones, related to the spin of the black hole.   

These new  rotation laws would allow   the investigation  of self-gravitating fluid bodies in the regime of strong gravity for general-relativistic versions of non-Keplerian rotation curves. 
This is not a subject of theoretical interest only --- there exists
a number of active galactic nuclei, that are known to posses  gaseous disks or toroids in their centers, and   that move according to non-Keplerian rotation laws  \cite{BL,KGM,KGM1,Combes}. Our rotation curves might be helpful in modelling these systems.

  Approximately stationary    disks can exist in tight accretion systems with central black holes. These highly relativistic systems  can be created in the merger of compact binaries consisting of pairs of black holes and neutron stars  \cite{Pan_Ton_Rez, Lovelace,MS}.  In the case of light disks --- with disks's masses much smaller than the mass of a black hole --- their motion  is quite likely ruled by the general-relativistic Keplerian rotation law \cite{KKMMOP,KKMMOP1}.

\section{Appendix: the conformal Kerr metric in 1PN approximation}
 
In this work we use  metrics that have conformally flat spatial parts in the 1PN order.  The
Kerr solution in standard coordinates  --- Boyer-Lindquist or Kerr-Schild --- does not belong to this class.
Fortunately, the Kerr geometry in harmonic coordinates \cite{Kerr_harmonic_coordinates} possesses a  conformally flat spatial metric in the 1PN approximation. For that reason we use it in this paper.

The Kerr metric in harmonic coordinates takes the following form \cite{Kerr_harmonic_coordinates}
\begin{multline}
ds^{2}=-c^{2}dt^{2}+\frac{R^{2}\left(R+\frac{GM}{c^{2}}\right)^{2}+a^{2}z^{2}}{\left(R^{2}+\frac{a^{2}}{R^{2}}z^{2}\right)^{2}}\times \\\left[\frac{\left(X^{i}dX_{i}+\frac{a^{2}}{R^{2}}zdz\right)^{2}}{R^{2}+a^{2}-\frac{G^{2}M^{2}}{c^{4}}} +\frac{z^{2}}{R^{2}}\frac{\left(X^{i}dX_{i}-\frac{R^{2}}{z}dz\right)^{2}}{R^{2}-z^{2}}\right]+\\
+\frac{2\frac{GM}{c^{2}}\left(R+\frac{GM}{c^{2}}\right)}{\left(R+\frac{GM}{c^{2}}\right)^{2}+\frac{a^{2}}{R^{2}}z^{2}}\times \\ \left[\frac{R\frac{G^{2}M^{2}}{c^{4}}a^{2}\left(R^{2}-z^{2}\right)\left(X^{i}dX_{i}+\frac{a^{2}}{R^{2}}zdz\right)}{\left(R^{2}+a^{2}-\frac{G^{2}M^{2}}{c^{4}}\right)\left(R^{2}+a^{2}\right)\left(R^{4}+a^{2}z^{2}\right)}+\right.\\
\left.+\frac{a\left(ydx-xdy\right)}{R^{2}+a^{2}}+cdt\right]^{2}+\\
\frac{\left(R+\frac{GM}{c^{2}}\right)^{2}+a^{2}}{R^{2}-z^{2}}\left[\frac{R^{2}\frac{G^{2}M^{2}}{c^{4}}a\left(R^{2}-z^{2}\right)\left(X^{i}dX_{i}+\frac{a^{2}}{R^{2}}zdz\right)}{\left(R^{2}+a^{2}-\frac{G^{2}M^{2}}{c^{4}}\right)\left(R^{2}+a^{2}\right)\left(R^{4}+a^{2}z^{2}\right)}\right.\\
\left.+\frac{R\left(ydx-xdy\right)}{R^{2}+a^{2}}\right]^{2},\label{Kerr_metric_harmonic}
\end{multline}
where $\left(X^{i}\right)=\left(x,y,z\right)$, $\left(dX_{i}\right)=\left(dx,dy,dz\right)$
and $R$ is given by the following equation
\[
\frac{x^{2}+y^{2}}{R^{2}+a^{2}}+\frac{z^{2}}{R^{2}}=1.
\]

The metric (\ref{Kerr_metric_harmonic})  in the 1PN approximation reads, in cylindrical coordinates ($x=r\cos\phi$,
$y=r\sin\phi$), 
\begin{multline}
ds^{2}=-\left[1+\frac{2U_{0}}{c^{2}}+\frac{2}{c^{4}}\left(U_{0}^{2}+U_{1}\right)\right]c^{2}dt^{2}-\frac{2}{c^{2}}A_{\phi}dtd\phi+\\
+\left(1-\frac{2U_{0}}{c^{2}}\right)\left(dr^{2}+dz^{2}+r^{2}d\phi^{2}\right).\label{Kerr_harmonic_1PN}
\end{multline}
Here
\begin{eqnarray}
U_{0} & = & -\frac{GM}{\sqrt{r^{2}+z^{2}}}\nonumber \\
U_{1} & = & -\frac{GJ^{2}\left(r^{2}-2z^{2}\right)}{2M\left(r^{2}+z^{2}\right)^{5/2}}\nonumber \\
A_{\phi} & = & \frac{2GJr^{2}}{\left(r^{2}+z^{2}\right)^{3/2}}.
\label{afi}
\end{eqnarray}

The circumferential radius $r_{\textrm{c}}$ is obtained from the $g_{\phi\phi}$
component of the metric (\ref{Kerr_harmonic_1PN})
\[
r_{\textrm{c}}=r\left(1-\frac{U_{0}}{c^{2}} \right).
\]

One can check that the angular velocity $\Omega$ of dust particles takes the following form at the equatorial plane (again up to the  1PN approximation),
\[
\Omega=\Omega_{0}+\frac{\Omega_{1}}{c^{2}} ,
\]
where:
\begin{eqnarray}
\label{aomega1}
\Omega_{0} & = & \frac{\sqrt{GM}}{r_{\textrm{c}}^{3/2}},\nonumber\\
\Omega_{1} & = & \frac{3J^{2}\sqrt{G}}{4M^{3/2}r_{\textrm{c}}^{7/2}}-\frac{GJ}{r_{\textrm{c}}^{3}}.
\end{eqnarray}

\begin{acknowledgments}
 We thank Patryk Mach for reading the manuscript and useful remarks.
\end{acknowledgments}

\end{document}